\documentstyle[12pt]{article}
\begin{document}
\pagestyle{empty}
%
% ----- VARIOUS DEFINITIONS
%
\def\id{{\bf 1}}
\def\0{{\emptyset}}
\def\sconf{{\underline{\sigma}}}
\def\hconf{{\underline{\chi}}}
\renewcommand{\theequation}{\arabic{section}.\arabic{equation}}
%
%---- EQUATION-COUNTER SET TO 0 WHEN A NEW SECTION BEGINS
%
\catcode `\@=11
%
%---- SHORT-COMMANDS
%
\def\be{\begin{equation}}
\def\ee{\end{equation}}
\newcommand{\ba}{\begin{eqnarray}}
\newcommand{\ea}{\end {eqnarray}}
\newcommand{\nn}{\nonumber}
\def\ri{\right}
\def\le{\left}
\def\mb{\makebox}
\def\mbo{\makebox(0,0)}
\def\mput{\multiput}
%
%---- GREEK-LETTERS
%
\def\a{\alpha}
\def\b{\beta}
\def\g{\gamma}
\def\G{\Gamma}
\def\d{\delta}
\def\D{\Delta}
\def\eps{\epsilon}
\def\vareps{\varepsilon}
\def\e{\eta}
\def\th{\theta}
\def\k{\kappa}
\def\l{\lambda}
\def\L{\Lambda}
\def\m{\mu}
\def\n{\nu}
\def\rh{\rho}
\def\p{\phi}
\def\s{\sigma}
\def\sp{\sigma^+}
\def\sm{\sigma^-}
\def\sx{\sigma^x}
\def\sy{\sigma^y}
\def\sz{\sigma^z}
\def\t{\tau}
\def\Th{\Theta}
\def\z{\zeta}
\def\vf{\varphi}
\def\vfi{\varphi^{(i)}}
\def\c{\chi}
\def\O{\Omega}
\def\Oh{\widehat\Omega}
\def\w{\omega}
%
%---- SYMBOLS
%
\def\ap{a^+}
\def\cp{c^+}
\def\woo{w_{0,0}}
\def\woi{w_{0,1}}
\def\wio{w_{1,0}}
\def\wii{w_{1,1}}
\def\Eoo{E^{00}}
\def\Eoi{E^{01}}
\def\Eio{E^{10}}
\def\Eii{E^{11}}
\def\Et{\tilde{E}}
\def\nil{\emptyset}
\def\i{\int}
\def\np{\not p}
\def\o{\over}
\def\pr{\prime}
\def\ra{\rightarrow}
\def\lra{\longrightarrow}
\def\ti{\tilde}
\def\v{\vert}
\def\inf{\infty}
%
%---- OPERATORS
%
\def\X{\times}
\def\ten{\otimes}
\def\grad{\nabla}
\def\pa{\partial}
%
%---- FRACTIONS ----
%
\def\half{\frac{1}{2}}
\def\quart{\frac{1}{4}}
%
%
%
%
%----------------------------------------------------------------------
% Titel page:
%----------------------------------------------------------------------
%
\ \\[12mm]
\begin{center}
	{\bf SOLUTION OF A ONE-DIMENSIONAL DIFFUSION-REACTION MODEL\\
	     WITH SPATIAL ASYMMETRY
         \footnote{Submitted to Zeitschrift f\"{u}r Physik}}
	     \\[20mm]
\end{center}
\begin{center}
\normalsize
	Haye Hinrichsen$^{\diamond}$
	% Hayes new address:
	\footnote{Present address :
	Dept. of Physics of Complex Systems,
	Weizmann Institute of Science, Rehovot 76100, Israel}
	, Klaus Krebs$^{\star}$
	 and Ingo Peschel$^{\diamond}$ \\[13mm]
	$^{\diamond}$ {\it Freie Universit\"{a}t Berlin, Fachbereich Physik\\
	 Arnimallee 14, D-14195 Berlin, Germany}\\[4mm]
	$^{\star}$ {\it Universit\"{a}t Bonn,
	Physikalisches Institut \\ Nu\ss allee 12,
   D-53115 Bonn, Germany}
\end{center}
\vspace{1cm}
%	\begin{center}
%
%	\end{center}
	\vspace{2.5cm}
{\bf Abstract:}
We study classical particles on the sites of an open chain which
diffuse, coagulate and decoagulate preferentially in one direction.
The master equation is expressed in terms of a spin one-half
Hamiltonian $H$ and the model is shown to be completely solvable
if all processes have the same asymmetry.
The relaxational spectrum is obtained directly from $H$ and via
the equations of motion for strings of empty sites.
The structure and the solvability of these equations are
investigated in the general case. Two phases are shown to exist
for small and large asymmetry, respectively,
which differ in their stationary properties.
\\[12mm]
\rule{6.6cm}{0.2mm}
\begin{flushleft}
\parbox[t]{3.5cm}{\bf Key words:}
\parbox[t]{12.5cm}{Reaction-diffusion systems,
		   non-equilibrium statistical mechanics,\\
		   coagulation model, integrability, spatial asymmetry}
\\[2mm]
\parbox[t]{3.5cm}{\bf PACS numbers:}
\parbox[t]{12.5cm}{05.40.+j, 05.70.Ln, 82.20.Mj}
\end{flushleft}
\normalsize
\thispagestyle{empty}
\mbox{}
%
%
%----------------------------------------------------------------------
% 1) INTRODUCTION:
%----------------------------------------------------------------------
%
\newpage
\setcounter{page}{1}
\setcounter{equation}{0}
\pagestyle{plain}
\section{Introduction}

  The study of exactly solvable models has been an important
activity in equilibrium statistical physics and it is equally
desirable to find such models in the field of non-equilibrium
phenomena. In this context
it has proved very useful to exploit the formal analogy between
the master equation for classical stochastic systems and
the quantum-mechanical Schr\"odinger equation. For example,
by relating the one-dimensional kinetic Ising model to a quantum
spin chain \cite{Felderhof,Siggia,Kimball,Peschel} it becomes
obvious why the system is solvable with
Glauber's \cite{Glauber} choice of the rates.
Similarly, the hopping of
classical particles with hard-core repulsion on a lattice
can be formulated as a spin problem, namely the ferromagnetic
Heisenberg model \cite{Alexander,Dieterich},
and in one dimension this is again a
solvable model.
\\ \indent
In recent work, this approach has been extended to more general
situations. They include diffusion with a preferred direction
as well as reactions between particles
or coupling to external reservoirs
\cite{Lushnikov,Alcaraz1,Alcaraz2,SandowSchuetz,GwaSpohn,Henkel,Stinchcombe}.
The spin of the resulting quantum models is $s+1$
if $s$ types of particles are involved. Fermionic or bosonic
representations have also been used, for example for models of
self-organized criticality \cite{Trimper,Cardy}.
In general, the inclusion
of more processes makes the time evolution operator more
complicated. On the other hand, by tuning the various rates one
may also achieve simplifications which make the problem integrable
in some cases.
\\ \indent
One of the systems where this happens is the coagulation
model in which  particles hop on a one-dimensional lattice
and, in addition, can merge and separate again with certain
probabilities. This model was introduced and treated in a continuum
approximation by Doering et.al.
\cite{Doering,Burschka,ben-Avraham,Horsthemke}.
It was found later that it is exactly solvable on a lattice if
the rates for hopping and for coagulation are equal \cite{Hinrichsen}.
The time evolution operator can then be brought
into the form of the Hamiltonian for a spin one-half $XY$ chain with a
$Z$ field. It is the dual of the operator for the Glauber model
mentioned above and can be diagonalized in terms of fermions.
The spectrum has a gap determined by the decoagulation rate.
If this rate vanishes, one finds an algebraic decay of the
concentration with exponent minus one-half. The model can then
be mapped onto a pair annihilation model \cite{Hinrichsen} and
there is also an interesting physical realization in terms of
excitons in the quasi one-dimensional substance TMMC \cite{Kroon}.
\\ \indent
In treating this model, it turned out that physical quantities
like the density are most easily calculated
from the probabilities to find empty intervals of arbitrary
length ("holes"). Their equations of motion form closed sets and
can be solved for a ring as well as for a chain.
They have been used also in related models \cite{Privman}.
An intriguing
aspect is that these closed sets even exist if the creation of
particles is allowed \cite{PRS}.
The relaxational spectrum of the hole
functions then has the form of the Wannier-Stark ladder found for
lattice electrons in a homogeneous electric field \cite{Stey,Saitoh}.
However, no fermionic solution is possible in this case and the
question of complete integrability has remained open so far.
\\ \indent
In the present article we study the properties of the coagulation
model for the case of a preferred direction. The geometry will
be the open chain where an interplay between the boundaries
and the spatial asymmetry occurs.
Our aim is to pursue the question of integrability further,
and also to find the basic physical features of the model.
We focus on the case where all processes have the same asymmetry
and the rates for hopping and coagulation are chosen equal.
Then the model contains two free parameters.
We show that, as for symmetric rates, the spin operator describing the
time evolution can be brought into a form where it is quadratic in fermions.
Thus one finds complete integrability also in this case.
However, in contrast to simple hopping or symmetric coagulation
the problem has no quantum group symmetry. This can be related to
particular boundary terms which correspond to fields in the magnetic
language and lead to special selection rules. There is, however,
a discrete $S_3$-symmetry which shows up in the multiplicity of
the relaxational modes.
\\ \indent
We also consider the hole equations, both in the one-hole sector
and in general. Following an approach by Bedeaux et.al. \cite{Bedeaux}
the different levels in the hierarchy can be decoupled and a com\-plete
solution achieved. By considering more general rates and
processes, it is seen that this procedure only works in the
free fermion case. This is also what the Reshetikhin criter\-ion
for integrability \cite{Reshetikhin} gives. Thus the other
cases where the
one-hole equations can be solved,  correspond to partially
integrable systems. Such a situation occurs also
in kinetic spin models \cite{Kimball,Deker} and in reaction-diffusion
models
where for particular rates the equations for the density correlations
form closed sets \cite{Schuetz}.
\\ \indent
The physical properties of the model are also interesting.
In the stationary state the density is inhomogeneous
and determined by a competition between asymmetry and decoagulation.
If the latter is small, the density is finite only near
one boundary. In the opposite case, it is constant in the bulk
of the system with some additional boundary effects. In between,
a transition takes place, where the profile becomes linear over
the major part of the system. These different regimes
are also reflected in the form of
the relaxation spectrum. In general, one finds a gap but at the
transition it vanishes and one has slow modes as in symmetric
diffusion. The phenomenon may be viewed as another example of
a boundary-induced transition
\cite{Henkel,Krug,DerridaDomany,SchuetzDomany}.
\\ \indent
We shall present the material as follows. In section 2 we introduce
the model and the spin formalism used in its treatment. The time
evolution operator is derived and brought into convenient forms in
section 3. In section 4 its spectrum is found by diagonalization via
fermions. Section 5 deals with the equations of motion in the
one-hole sector and in section 6 the $n$-hole problem is treated
via cumulant-like functions. From these results, density profile
and correlation functions in the steady state are obtained as special
cases and discussed in section 7. Finally, section 8  contains
a summary and concluding remarks.
%
%----------------------------------------------------------------------
% 2) DEFINITION OF THE MODEL:
%----------------------------------------------------------------------
%
\section{Model}
\setcounter{equation}{0}
We consider classical particles on a chain of $L$ sites, each of
which can either be empty or occupied with at most one particle.
The dynamics of the system
consists of transitions which involve only the configurations of
neighbouring sites. As discussed previously \cite{PRS}, one can
then distinguish altogether $12$ different processes, namely hopping,
coagulation and decoagulation as well as birth and death of single
particles and of pairs. Each process is described by a rate constant
and, in the absence of detailed balance, all these constants are
independent.
\\ \indent
In the bulk of the paper we will only consider the first three
processes which are shown below together with their rates:
%
%
% Pictures describing the processes:
%
%
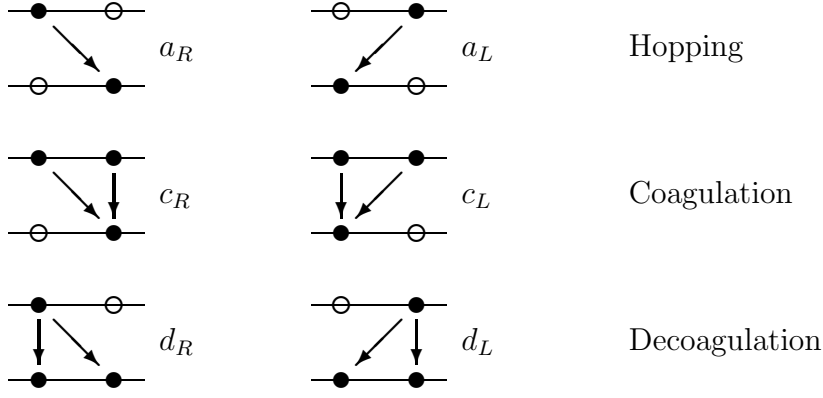
\def\latt{\thinlines\put(0,1){\line(1,0){9}}
			 \put(0,6){\line(1,0){9}}\thicklines}
\def\arr{\put(3,5){\vector(1,-1){3}}}
\def\arl{\put(6,5){\vector(-1,-1){3}}}
\def\dol{\put(2,5){\vector(0,-1){3}}}
\def\dor{\put(7,5){\vector(0,-1){3}}}
\def\occ{\circle*{1}}
\def\vac{\circle{1}}
\def\picar{\begin{picture}(18,7)\latt \arr
\put(2,6){\occ} \put(7,6){\vac} \put(10,3){$a_R$}
\put(2,1){\vac} \put(7,1){\occ} \end{picture}}
\def\pical{\begin{picture}(18,7)\latt \arl
\put(2,6){\vac} \put(7,6){\occ} \put(10,3){$a_L$}
\put(2,1){\occ} \put(7,1){\vac} \end{picture}}
\def\piccr{\begin{picture}(18,9)\latt \arr \dor
\put(2,6){\occ} \put(7,6){\occ} \put(10,3){$c_R$}
\put(2,1){\vac} \put(7,1){\occ} \end{picture}}
\def\piccl{\begin{picture}(18,9)\latt \arl \dol
\put(2,6){\occ} \put(7,6){\occ} \put(10,3){$c_L$}
\put(2,1){\occ} \put(7,1){\vac} \end{picture}}
\def\picdr{\begin{picture}(18,9)\latt \arr \dol
\put(2,6){\occ} \put(7,6){\vac} \put(10,3){$d_R$}
\put(2,1){\occ} \put(7,1){\occ} \end{picture}}
\def\picdl{\begin{picture}(18,9)\latt \arl \dor
\put(2,6){\vac} \put(7,6){\occ} \put(10,3){$d_L$}
\put(2,1){\occ} \put(7,1){\occ} \end{picture}}
\def\pichopp{\begin{picture}(18,7)\put(1,3){Hopping}\end{picture}}
\def\piccoag{\begin{picture}(18,7)\put(1,3){Coagulation}\end{picture}}
\def\picdeco{\begin{picture}(18,7)\put(1,3){Decoagulation}\end{picture}}

\begin{center}
\setlength{\unitlength}{2mm}
\begin{tabular}{ccl}
\picar & \pical & \pichopp \\
\piccr & \piccl & \piccoag \\
\picdr & \picdl & \picdeco
\end{tabular}
\end{center}

The left-right asymmetry which we assume could result from external
fields. We will use a spin language to describe the system. The
state of site $n$ is specified by a variable $\sigma_n=\pm 1$ such that
$\sigma_n=+1(-1)$ if the site is occupied (empty). A configuration
of the total system is denoted by $\sconf=\{\sigma_1,\sigma_2,\ldots
\sigma_L\}$. The probability $P(\sconf,t)$ to find configuration
$\sconf$ at time $t$ then obeys the master equation
\begin{equation}
\frac{\partial}{\partial t} \, P(\sconf,t) \;=\;
-\sum_{\sconf'} \, P(\sconf,t) \, w(\sconf \rightarrow \sconf') +
\sum_{\sconf'} \, P(\sconf',t) \, w(\sconf' \rightarrow \sconf)
\end{equation}
\noindent
where the two terms on the right represent the loss and the gain
processes, respectively. The transition probabilities
$w(\sconf \rightarrow \sconf')$ are sums of contributions from the
various nearest neighbour processes.
\\ \indent
To formulate the problem in quantum-mechanical terms
\cite{Felderhof,Kadanoff} one uses a ket vector
notation in the $2^L$-dimensional state space, writing
\begin{equation}
|P(t)\rangle \;=\; \sum_\sconf \, P(\sconf,t)\,|\sconf\rangle\,.
\end{equation}
\noindent
The master equation then takes the form
\begin{equation}
\label{EqOfMotion}
\frac{\partial}{\partial t}\,|P(t)\rangle \;=\;
-H |P(t)\rangle
\end{equation}
\noindent
from which the analogy with the Schr\"odinger equation is obvious.
The time evolution operator $H$ will therefore be called
Hamiltonian in the following. However, since $|P(t)\rangle$ is
already the probability, expectation values differ from the
quantum-mechanical ones. For a quantity $A(\sconf)$ one has
\begin{equation}
\langle A(t) \rangle \;=\;
\sum_\sconf \, A(\sconf) \, P(\sconf,t) \;=\;
\langle 0 | A | P(t) \rangle
\end{equation}
\noindent
with the bra vector
\begin{equation}
\langle 0 | \;=\; \sum_\sconf \, \langle \sconf |\,.
\end{equation}
\noindent
The operator $H$ is the sum of nearest-neighbour terms
\begin{equation}
\label{chain}
H \;=\; \sum_{n=1}^{L-1} \, H_{n,n+1}
\end{equation}
\noindent
and the $(4 \times 4)$ matrix $H_{n,n+1}$ can be written down
easily \cite{PRS}. In a basis of states
$(\,|--\rangle,\,|-+\rangle,\,|+-\rangle,\,|++\rangle)$
the result is, including birth processes (rates $b_R, b_L$)
and pair creation (rate $c$) for further reference
\begin{equation}
\label{Hamiltonian}
H_{n,n+1} \;=\; \left(\begin{array}{cccc}
			c+b_L+b_R & 0 & 0 & 0 \\
		  -b_R & a_L+d_L & -a_R & -c_R \\
		  -b_L & -a_L & a_R+d_R & -c_L \\
		  -c   & -d_L & -d_R & c_L+c_R
		  \end{array} \right)\,.
\end{equation}
\noindent
Due to probability conservation, the elements in the columns
sum to zero. $H_{n,n+1}$ and therefore the total operator $H$
is non-hermitean.
\\ \indent
In the next sections we will consider the case of no birth and
pair creation $(b_R=b_L=c=0)$, equal rates for hopping and
coagulation $(c_R=a_R, c_L=a_L)$ and equal asymmetry for all
processes. Up to a common constant which we set equal one,
the rates can then be written
\begin{equation}
\label{rates}
a_R=\frac{1}{q}\,,
\hspace{6mm}
a_L=q\,,
\hspace{6mm}
d_R=\frac{\Delta}{q}\,,
\hspace{6mm}
d_L=\Delta q\,.
\end{equation}
\noindent
Thus the parameter $q$ as usual describes the asymmetry, while
$\Delta$ gives the ratio of decoagulation to coagulation rates and
therefore governs the (average) density.
%
%
%----------------------------------------------------------------------
% 3) TIME EVOLUTION OPERATOR
%----------------------------------------------------------------------
%
\section{Time evolution operator}
\setcounter{equation}{0}
Starting e.g. from Eq. (\ref{Hamiltonian}), the Hamiltonian $H$
can be expressed in terms of Pauli-matrices
$\sigma_i^\alpha \; (\alpha$=$x,y,z)$. It is the sum of diffusion
and coagulation contributions
\begin{equation}
H \;=\; H_D + H_C\,,
\end{equation}
\noindent
where $H_D$ is given by
\begin{eqnarray}
H_D &=& -\frac14 \, \sum_{i=1}^{L-1}
\biggl\{
q^{-1}\sigma_i^-\sigma_{i+1}^+ + q \,\sigma^+_i\sigma_{i+1}^-
+ (q+q^{-1})\,\sigma_i^z\sigma_{i+1}^z \\
&& \hspace{16mm}
+(q-q^{-1}) (\sigma_i^z-\sigma_{i+1}^z) - (q+q^{-1})
\biggr\}\,. \nonumber
\end{eqnarray}
\noindent
The standard $U_q[SU(2)]$-symmetric form of $H_D$ is then obtained
by a position-dependent rescaling of $\sigma^+$ and $\sigma^-$
\cite{Alcaraz1,SandowSchuetz,Henkel}.
Physically, this corresponds to factorizing out of
$P(\sconf,t)$ the quantity $P_0^{1/2}(\sconf)$, where $P_0(\sconf)$
is the stationary distribution
in the case of simple hopping. The coagulation part reads
\begin{eqnarray}
H_C &=& -\frac14 \, \sum_{i=1}^{L-1}
\biggl\{
(q^{-1}\sigma_i^-+\Delta\,q\,\sigma_i^+)(1+\sigma_{i+1}^z) \,+\,
(1+\sigma_i^z)(q \sigma_{i+1}^- + \Delta q^{-1} \sigma_{i+1}^+)
\nonumber \\ && \hspace{16mm}
- \, (q+q^{-1}) (1-\Delta)\,\sigma_i^z\sigma_{i+1}^z \,-\,
(q+q^{-1})(\sigma_i^z+\sigma_{i+1}^z)
\\ && \hspace{16mm}
+ \, \Delta\, (q-q^{-1})(\sigma_i^z-\sigma_{i+1}^z) \,-\,
(q+q^{-1})(1+\Delta)
\nonumber \biggr\}   \,.
\end{eqnarray}
\noindent
It is more complicated than $H_D$ because of the single spin terms
$\sigma_i^+$, $\sigma_i^-$, $\sigma_i^z$ which appear in it.
One should note, however, that the contribution
$(\sigma_i^z-\sigma_{i+1}^z)$ is effectively only a boundary term.
For $q=1$ it vanishes altogether and $H$ reduces to the expression
found previously \cite{Alcaraz1,Hinrichsen}.
In that case, $H$ was brought into a
simpler form by two local transformations in spin space:
A hyperbolic rotation around the $z$-axis with generator
\begin{equation}
V_1=\prod_n\,e^{\lambda \sigma_n^z}\,;
\hspace{15mm}
e^{-2\lambda} = \sqrt{\Delta}
\end{equation}
\noindent
which rescales the $\sigma^\pm$ operators, and a real rotation around
the $y$-axis with generator
\begin{equation}
V_2=\prod_n\,e^{-i \frac{\Theta}{2} \sigma_n^y}\,;
\hspace{15mm}
cos(\Theta) = \frac{1}{\sqrt{1+\Delta}}
\end{equation}
\noindent
which forms linear contributions of $\sigma^x$ and $\sigma^z$.
\\ \indent
It turns out that the same transformation can be used also for
$q \neq 1$. The result for the new operator $H'=V_2V_1HV_1^{-1}V_2^{-1}$
then is
\begin{eqnarray}
H' &=& -\frac{D}{2} \, \sum_{i=1}^{L-1} \,
\biggl\{
\gamma\,\sigma_i^x\sigma_{i+1}^x \,+\,
\gamma^{-1}\sigma_i^y \sigma_{i+1}^y \,+\,
i\,\frac{1-q^2}{1+q^2}\,(\sigma_i^x\sigma_{i+1}^y-\sigma_i^y\sigma_{i+1}^x)
\nonumber \\ && \hspace{16mm}
-\, \frac{2}{q+q^{-1}} \, (q^{-1}\sigma^z_i + q \sigma_{i+1}^z)
\,-\, (\gamma+\gamma^{-1})
\\ && \hspace{16mm}
-\,\frac{1-q^2}{1+q^2}\,\sqrt{\gamma-\gamma^{-1}}\,
\biggl[ \sqrt{\gamma} \,(\sigma_i^x-\sigma_{i+1}^x) \,+\,
i\,\frac{1}{\sqrt{\gamma}}\,(\sigma_i^y-\sigma_{i+1}^y)\biggr] \biggr\}
\nonumber
\end{eqnarray}
\noindent
where $\gamma=\sqrt{1+\Delta}$.
and $D=\frac12(q+q^{-1})\gamma$.
\\ \indent
This operator has two non-hermitean parts: the third term, which is
a kind of chiral contribution from the bulk, and the last term
which again acts only at the two boundaries. Both vanish in the
symmetric case $(q=1)$ and $H'$ then reduces to the dual of the
Glauber model operator.
\\ \indent
The bulk terms in $H'$ are all quadratic in fermion operators after
a Jordan-Wigner transformation. There are, however, also boundary
terms in $\sigma^x$, $\sigma^y$ and $\sigma^z$. The last one is
unproblematic, but $\sigma^x$ and $\sigma^y$ are linear in the fermions.
There is, however, a well-known way around this difficulty: One
extends the chain by one site at each end and uses the additional spin
operators, which are constants of the motion, to make boundary
terms quadratic. Thus the operator $H'$ is solvable via fermions.
\\ \indent
Actually, one can bring it into a slightly simpler form by performing
a rotation by $180^\circ$
\begin{equation}
V_3=\prod_n\,e^{i\frac{\pi}{2}\sigma_n^y}
\end{equation}
and a further transformation with the generator
\begin{equation}
V_4=\prod_n\,e^{\rho \sigma_n^z}\,;
\hspace{15mm}
e^{-2\rho} = \sqrt{\frac{\gamma-1}{\gamma+1}}
\end{equation}
\noindent
to obtain
\begin{eqnarray}
\label{StandardForm}
H'' &=& -\frac12 \, \gamma \, \sum_{i=1}^{L-1} \,
\biggl\{ (\eta+\eta^{-1}) \sigma^x_i \sigma_{i+1}^x +
i\,(q \sigma_i^x \sigma_{i+1}^y + q^{-1} \sigma_i^y\sigma_{i+1}^x)
\\ && \hspace{20mm}
+ (q^{-1} \sigma_i^z+q \sigma_{i+1}^z) -
\frac{1}{2} (q-q^{-1}) (\gamma-\gamma^{-1})
  (\sigma_i^x-\sigma_{i+1}^x) - (\eta+\eta^{-1}) \biggr\}
\nonumber
\end{eqnarray}
\noindent
where $(\eta+\eta^{-1}) = \frac12(q+q^{-1})(\gamma+\gamma^{-1})$.
\\ \indent
There is no $\sigma^y$ boundary term in this expression so that the
only non-hermitean part is the chiral bulk term. This operator will
be diagonalized in the next section.
\\ \indent
We mention that $H'$ and $H''$ have a simple symmetry, namely
they are invariant under the tranformation $q\leftrightarrow 1/q$
combined with spatial reflection $i\rightarrow L-i+1$.
%
%
%
%----------------------------------------------------------------------
% 4) ANALYSIS OF THE SPECTRUM
%----------------------------------------------------------------------
%
\section{Analysis of the spectrum}
\setcounter{equation}{0}
In order to diagonalize the Hamiltonian (\ref{StandardForm})
in terms of fermions, we write $H''=H_0''+H_1''$ where
\begin{equation}
H_1'' \;=\; \frac{\kappa}{2} (\sigma_1^x-\sigma_{L}^x)\,
\end{equation}
\noindent
is the linear boundary term and
\begin{equation}
\label{Kappa}
\kappa \;=\; \frac{\gamma}{2} (q-q^{-1})(\gamma-\gamma^{-1})\,.
\end{equation}
\noindent
As outlined in the last section, one appends
one site at each end of the chain and extends
$H_1''$ by two Pauli matrices $\sigma_0^x$ and
$\sigma_{L+1}^x$. Then
\begin{equation}
\tilde{H}_1'' \;=\; \frac{\kappa}{2}
(\sigma_0^x\sigma_1^x-\sigma_{L}^x\sigma_{L+1}^x)\,
\end{equation}
\noindent
is bilinear in fermions so that standard diagonalization methods
\cite{LSM} can be applied\footnote{In the same way one could
also add $\sigma^y$-matrices or any linear combination of
$\sigma^x$ and $\sigma^y$.}. The eigenvectors
of the extended Hamiltonian $\tilde{H}''=H_0''+\tilde{H}_1''$
include those of the original chain $H''$ as follows:
Since the appended matrices are constants of the motion,
the spectrum of $\tilde{H}''$ decomposes into four sectors
$(++,+-,-+,--)$ corresponding to the eigenvalues of $\sigma_0^x$
and $\sigma_{L+1}^x$. This implies that all eigenvectors of $\tilde{H}''$
have the form
\begin{equation}
|\tilde{\psi}\rangle \;=\;
\left( \begin{array}{c} 1 \\ \pm 1 \end{array} \right)
\otimes |\psi\rangle \otimes
\left( \begin{array}{c} 1 \\ \pm 1 \end{array} \right)\,,
\end{equation}
\noindent
where $|\psi\rangle$ is a vector with $2^L$ components. Since in each
sector the appended matrices can be replaced by their eigenvalues,
the spectrum of the original chain appears in
the $(++)$ sector and the vectors $|\psi\rangle$
are just the eigenvectors of $H''$.
Therefore we can determine the eigenvectors of the original
problem by projection onto the $(++)$ sector. One should mention
that a very similar situation occurs in calculations using the
corner transfer matrix of the two-dimensional Ising model
\cite{Baxter,Igloi}. In that case the sectors arise from the fixed inner
and outer spins in the corner geometry.
\\ \indent
Defining anticommuting Clifford operators $\tau_j^{x,y}=
(\prod_{i=0}^{j-1}\sigma_i^z)\sigma_j^{x,y}$ one can rewrite $\tilde{H}''$
as a bilinear expression in $\tau_j^x$ and $\tau_j^y$. Then a second
linear transformation
$a_k^\pm=\frac12\sum_{j=1}^L (\phi_{k,j}^x \tau_j^x \pm
i \phi_{k,j}^y \tau_j^y)$ yields
\begin{equation}
\label{DiagonalForm}
\tilde{H}'' \;=\; \Big( \sum_k \lambda_k \, a^+_ka^-_k \Big) \,-\,\kappa
\end{equation}
\noindent
which is a diagonal expression in fermionic
creation and annihilation operators $a_k^+$ and $a_k^-$.
The one-particle energies $\lambda_k$
and the eigenvectors $\vec{\phi}^\pm_k = (\phi_{k,1}^x,
\pm i \phi_{k,1}^y,\ldots,\phi_{k,L}^x,\pm i \phi_{k,L}^y)$
can be derived from the eigenvalue problem
$M\vec{\phi}^\pm_k=\mp\lambda_k \vec{\phi}^\pm_k$ with
\begin{equation}
\label{M}
M \,=\, i\gamma\,\left(
	 \begin{array}{ccccccc}
	 0 & D & & & & & \\
	 -D^T & B & A & & & & \\
	 & -A^{T} & B+C & A & & & \\
	 & & ... & ... & ... & & \\
	 & & & -A^{T} & B+C & A & \\
	 & & & & -A^{T} & C & -D \\
	 & & & & & D^T & 0
	 \end{array}
\right)
\end{equation}
\noindent
where $A, B, C$ and $D$ denote the $2 \times 2$ block matrices
\begin{eqnarray*}
A=\left( \begin{array}{cc} -i/q & 0 \\
\eta+\eta^{-1} & iq \end{array} \right)\,,
\hspace{5mm}
B=\left( \begin{array}{cc} 0 & q^{-1} \\ -q^{-1} & 0 \end{array} \right)\,,
&&\hspace{1mm}
C=\left( \begin{array}{cc} 0 & q \\ -q & 0 \end{array} \right)\,,
\hspace{5mm}
D=\left( \begin{array}{cc} 0 & 0 \\
\kappa/\gamma & 0 \end{array} \right).
\end{eqnarray*}
\noindent
The eigenvalues of $M$ are real and occur in pairs with different signs.
It should be emphasized at this point that although
$a^+_k$ and $a^-_k$ obey the usual fermionic commutation relations
$\{a_k^+,a_l^-\}=\delta_{kl}$, we have $a_k^+ \neq (a_k^-)^\dagger$
because of the nonhermiticity of $\tilde{H}''$.
\\ \indent
Assuming that the wave function $\phi^\pm_{k,j}$
consists of exponential contributions $e^{-\zeta j}$, the
bulk equations (\ref{M}) yield the dispersion relation
\begin{equation}
\label{Dispersion}
\lambda \;=\; \gamma\,
(\eta+\eta^{-1}-q e^{-\zeta}-q^{-1} e^{\zeta})\,.
\end{equation}
\noindent
Including the boundary conditions which determine $\zeta$,
we obtain three types of fermionic
one-particle states:
\begin{itemize}
\item One trivial zero mode ($\lambda_0=0$) which
		is localized at the ends of the extended chain.
		This zero mode represents conserved quantities
		$\sigma_0^x$ and $\sigma_{L+1}^x$
		and causes two-fold degeneracies of all levels.
\item Two exponential modes with excitation energies
		$\lambda_I=\lambda_{II}=\kappa$. Their wave functions
		$\phi_I^\pm$ and $\phi_{II}^\pm$ decay exponentially
		with three different length scales corresponding
                to the values
		$\ln(\gamma)$, $\ln(q^2\gamma)$ and $\ln(q^2\gamma^{-1})$
		for $\zeta$. These modes reflect the
		influence of the boundary term $H_1''$.
\item $L-1$ oscillatory modes with
		$\lambda_k=\gamma(\eta+\eta^{-1})-
		2\gamma\cos\frac{\pi k}{L}$ where $k=1,\ldots,L-1$.
		The corresponding wave functions show damped oscillations
		of the form	$\phi_{k,j} \sim q^{\pm j} e^{\pm i\pi k/L}$.
		This implies the presence of a fourth length
		scale with $\zeta=\ln{q}$. The damping is a
		consequence of the nonhermiticity of the Hamiltonian
                for $q\neq 1$.
\end{itemize}

\noindent
One can show that the lowest level of $\tilde{H}''$ is a two-fold
degenerate state in the sectors $(+-)$ and $(-+)$
with a negative eigenvalue $\Lambda=-\kappa$.
The lowest level in the $(++)$ sector with eigenvalue zero
is already an excited state.
The total number of excited modes in this sector has to be odd and
one can distinguish:

\begin{itemize}
\item States with an even number of excited oscillatory modes.
The levels are two-fold degenerate because either mode $I$
or mode $II$ can be occupied. This class contains the ground
state with $\Lambda=0$ and the lowest gap is
\begin{equation}
\label{Gap0}
\Delta \Lambda \;=\;\gamma (\eta+\frac{1}{\eta}-2)\,.
\end{equation}
\noindent
\item States with an odd number of excited oscillatory modes.
Here the modes $I$ and $II$ are either both occupied, which gives a
contribution ($+\kappa$) to $\Lambda$, or both empty, which gives
($-\kappa$), i.e. we have two bands of excitations which
are shifted symmetrically by $\pm \kappa$. Their lowest gaps are
\begin{equation}
\label{Gaps}
\Delta \Lambda(\kappa) = \gamma(\gamma q+\frac{1}{\gamma q}-2)\,,
\hspace{15mm}
\Delta \Lambda(-\kappa) = \gamma(\frac{\gamma}{q}+\frac{q}{\gamma}-2)\,.
\end{equation}
\end{itemize}
The nature of these three branches (singlets and doublets) suggests
a symmetry in the problem. This is indeed the case: From the
operators for the two exponential modes one can construct the
quantities
\begin{eqnarray}
\sigma = 1 - ( a_I^{+} a^{-}_I + a_{II}^{+} a^{-}_{II} )
           + (a_{I}^{+} a^{-}_{II} + a_{II}^{+} a^{-}_{I})\\
\tau = 1
 -\frac32 (a_I^{+} a^{-}_I - a_{II}^{+} a^{-}_{II})^2
 -\frac{\sqrt{3}}{2} (a_{I}^{+} a^{-}_{II} - a_{II}^{+} a^{-}_{I})\nn .
\end{eqnarray}
They satisfy the relations
\begin{equation}
\sigma^2 = 1, \quad \tau^3 = 1, \quad \sigma \tau \sigma = \tau^2
\end{equation}
which characterize the generators of the permutation group $S_3$
and commute with both $\tilde H''$ and $H''$.
Thus the levels
can be chosen as irreducible representations of $S_3$ which,
according to group theory \cite{Hamermesh},
consist of two types of singlets and one doublet.
The operator $\sigma$ has a simple meaning,
it interchanges the two degenerate fermion modes.
\\ \indent
The smallest gap is obtained from one of the expressions
(\ref{Gaps}) which differ by
a reflection $q \rightarrow q^{-1}$.
It vanishes if
$\gamma=q$ $(q>1)$ or $\gamma=q^{-1}$ $(q<1)$. One band of
excitations is then massless and algebraic long-time behaviour
is expected.
At this point a transition takes place between a low-density phase
(for $\gamma<q$) and a high-density phase (for $\gamma>q$). This
will be seen in section 7 from the steady-state properties
of the system. One should note, however, that the situation is
different from a
Pokrovskii-Talapov type of transition \cite{Pokrovskii}
since here the spectrum
has a gap on both sides $(\gamma \stackrel{\scriptstyle <}
{\scriptstyle >} q)$ and always varies quadratically
with the momentum $k$.
%
%
%----------------------------------------------------------------------
% HOLE FORMALISM
%----------------------------------------------------------------------
%
\section{Hole formalism}
\setcounter{equation}{0}
\noindent
Although the diagonalization in terms of fermions solves the model
in principle, another complementary approach is simpler for
practical purposes and gives additional insight. In this approach
the states of the system are described in terms of strings of empty sites
(holes) instead of spin configurations \cite{Doering}.
In this way one can distinguish different
sectors characterized by the number of holes.
The equations for the probabilities couple a
sector only to the one with lower order.
For the one-hole sector of the
coagulation model (\ref{Hamiltonian}) with free ends they were
already derived in Ref. \cite{PRS}.
Denoting by $\Omega(x,y,t)$ the probability to find the sites
$x+1,\ldots,y$ empty and using the abbreviations
\begin{eqnarray}
\label{Abbreviations}
\alpha_{L,R} &=& a_{L,R}+d_{L,R}-b_{L,R}-c \\
\gamma_{L,R} &=& a_L + a_R + d_{L,R} \nonumber \\
\delta &=& b_L+b_R+c \nonumber\,
\end{eqnarray}
\noindent
they read
\begin{itemize}
\item for holes which do not touch the boundaries
($0<x<y<L$):
\begin{eqnarray}
\label{FullEquations}
\frac{d}{dt}\Omega(x,y,t) &=&
\alpha_R\,\Omega(x-1,y,t) \,+\,
a_L\,\Omega(x+1,y,t) \\
&&+\;
a_R\,\Omega(x,y-1,t) \,+\, \nonumber
\alpha_L\,\Omega(x,y+1,t) \\
&&-\; (\gamma_L+\gamma_R+(y-x-1) \delta) \,\Omega(x,y,t) \nonumber
\end{eqnarray}
\noindent
\item for holes touching the left boundary
($0=x<y<L$):
\begin{eqnarray}
\frac{d}{dt}\Omega(0,y,t) &=&
a_R\,\Omega(0,y-1,t) \,+\,
\alpha_L\,\Omega(0,y+1,t) \\ && -\, \nonumber
(\gamma_L+(y-1) \delta)\, \Omega(0,y,t)
\end{eqnarray}
\noindent
\item for holes touching the right boundary
($0<x<y=L$):
\begin{eqnarray}
\frac{d}{dt}\Omega(x,L,t) &=&
\alpha_R \Omega(x-1,L,t) +
a_L \Omega(x+1,L,t) \\ && -\, \nonumber
(\gamma_R+(L-x-1) \delta)\, \Omega(x,L,t)
\end{eqnarray}
\noindent
\item for the hole extending over the whole chain
($x=0, y=L$):
\begin{equation}
\label{FullEquationsEnd}
\frac{d}{dt}\Omega(0,L,t) \;=\;
-(L-1) \delta \,\Omega(0,L,t)\,,
\end{equation}
\noindent
\end{itemize}
where we formally define $\Omega(x,x,t)=1$ according to \cite{PRS}.
This is the coupling to the lower sector
(no holes), which here leads to an {\it inhomogeneous} system.
The integrable case discussed in the preceeding sections
corresponds to
\begin{equation}
\label{IntegrableCase}
\alpha_{L,R} = \gamma^2 q^{\pm 1},
\hspace{4mm}
a_{L,R} = q^{\pm 1},
\hspace{4mm}
\gamma_{L,R} = q^{\mp 1} + \gamma^2 q^{\pm 1},
\hspace{4mm}
\delta = 0\,.
\end{equation}
\noindent
Introducing rescaled probabilities $\tilde{\Omega}(x,y,t)$ via
\begin{equation}
\label{Rescaling}
\tilde{\Omega}(x,y,t) \;=\;
\gamma^{y-x} q^{y+x} \, \Omega(x,y,t)
\end{equation}
\noindent
and separating the time dependence $exp(-\Lambda t)$,
Eqs. (\ref{FullEquations})--(\ref{FullEquationsEnd}) reduce to:
\begin{eqnarray}
\label{ScaledEquations}
\Bigl(\gamma(q+q^{-1})(\gamma+\gamma^{-1})-\Lambda \Bigr) \tilde{\Omega}(x,y)
&=& \gamma \Bigl(\tilde{\Omega}(x-1,y)+\tilde{\Omega}(x+1,y)
\\ && \hspace{3mm} \nonumber
+\tilde{\Omega}(x,y-1)+\tilde{\Omega}(x,y+1)\Bigr)\hspace{3mm}
\\[2mm]
\Bigl(q^{-1}+\gamma^2 q-\Lambda\Bigr)\, \tilde{\Omega}(0,y)
&=& \gamma\Bigl(\tilde{\Omega}(0,y-1)+\tilde{\Omega}(0,y+1)\Bigr)
\\[2mm]
\Bigl(q+\gamma^2 q^{-1}-\Lambda\Bigr) \,\tilde{\Omega}(x,L)
&=& \gamma\Bigl(\tilde{\Omega}(x+1,L)+\tilde{\Omega}(x-1,L)\Bigr)
\\[2mm]
\label{ScaledEquationsEnd}
\Lambda\, \tilde{\Omega}(0,L) &=& 0
\end{eqnarray}
\noindent
with the inhomogeneous boundary condition
$\tilde{\Omega}(x,x)\;=\; q^{2x}$.
These equations can be solved using similar
techniques as in Ref. \cite{Hinrichsen} which rely mainly on the
invariance of the bulk equation (\ref{ScaledEquations})
under reflections $x \leftrightarrow y$ and $x \leftrightarrow L-y$.
A trivial {\it inhomogeneous}
solution is the empty lattice, for which
$\tilde{\Omega}(x,y)=\gamma^{y-x} q^{y+x}$.
The {\it homogeneous} set of solutions
satisfies the condition $\tilde{\Omega}(x,x)=0$ and reads
\begin{eqnarray}
\label{PhiZeroMode}
\Phi_0(x,y) &=& h(L-x)g(y)-h(L-y)g(x)\\
\Lambda_0 &=& 0 \nonumber \\[3mm]
\Phi^{(L)}_k(x,y) &=& \sin\frac{\pi k x}{L} \, h(L-y) -
\sin\frac{\pi k y}{L} \, h(L-x)  \\
\Lambda^{(L)}_k &=& \gamma\,\Big( \gamma q+\gamma^{-1}q^{-1} -
2 \cos\frac{\pi k}{L} \Big) \nonumber \\[3mm]
\Phi^{(R)}_k(x,y) &=& \sin\frac{\pi k x}{L} \, g(y)-
\sin\frac{\pi k y}{L} \, g(x)\\
\Lambda^{(R)}_k &=& \gamma\,\Big( \gamma q^{-1}+\gamma^{-1}q -
2 \cos\frac{\pi k}{L} \Big) \nonumber \\[3mm]
\label{PhiTwoFermions}
\Phi_{k,l}(x,y) &=& \sin\frac{\pi k x}{L}\,\sin\frac{\pi l y}{L}\,-\,
\sin\frac{\pi k y}{L}\,\sin\frac{\pi l x}{L}  \\
\Lambda_{k,l} &=& \gamma\,\Big( (q+q^{-1})(\gamma+\gamma^{-1}) -2
(\cos\frac{\pi k}{L}+\cos\frac{\pi l}{L})\Big) \nonumber
\end{eqnarray}
\noindent
where $1 \leq k < l \leq L-1$ and
the functions $g$  and $h$ are
\begin{equation}
g(x)=\frac{(\gamma q)^x-(\gamma q)^{-x}}{(\gamma q)^L-(\gamma q)^{-L}}\,,
\hspace{15mm}
h(x)=\frac{(\gamma/q)^x-(\gamma/q)^{-x}}{(\gamma/q)^L-(\gamma/q)^{-L}}\,.
\end{equation}
\noindent
We can now compare the eigenvalues $\Lambda_0,\Lambda_k^{L,R}$
and $\Lambda_{k,l}$ with the spectrum computed in the last section.
As in the symmetric case \cite{Hinrichsen}, $\Lambda_0$ gives a
two-fold degenerate ground state while $\Lambda_k^{L,R}$ and
$\Lambda_{k,l}$ are seen to be one- and two-particle excitations.
However, in contrast to the symmetric case we have
$\Lambda_k^{L} \neq \Lambda_k^{R}$ since these levels are shifted by
$\pm \kappa$. Therefore the boundary effects of $\tilde{H}_2''$ do
not change the selection rules, they only modify the mass gaps
of the one-particle modes.
%
%
%
%----------------------------------------------------------------------
% BEDEAUX FUNCTIONS
%----------------------------------------------------------------------
%
\section{Probabilities for n holes}
\setcounter{equation}{0}
For a complete treatment of the system  one has to study
configurations with an arbitrary number of holes.
This was already discussed in Ref. \cite {PRS} and we will now
carry out such a study. The probability to find the strings of
sites $(x_1+1,\cdots,y_1),(x_2+1,\cdots,y_2),
\cdots,(x_n+1,\cdots,y_n)$  with $0\leq x_1<\cdots <y_n\leq L$
empty at time $t$ is denoted by $\O_n(x_1,y_1,\cdots,x_n,y_n,t)$
and will also be called $n$--hole function. If no hole extends
up to the boundaries,
the time evolution of $\O_n$ takes the form :
\ba
&&\frac d {dt} \O_n(x_1,y_1,\cdots,x_n,y_n,t)\;=\; \sum_{i=1}^{n}
\biggl[-\Bigl(\g_L+\g_R+(y_i-x_i-1)\d \Bigl)\O_n(x_1,y_1,\cdots,x_n,y_n,t)
\label{nloch}\nn\\
&&\hspace{30mm}
+\a_R\O_n(\cdots,x_i-1,y_i,\cdots,t)
+a_L\O_n(\cdots,x_i+1,y_i,\cdots,t)\nn\\
&&\hspace{30mm}
+a_R\O_n(\cdots,y_i-1,x_{i+1},\cdots,t)+
\a_L\O_n(\cdots,y_i+1,x_{i+1},\cdots,t)
\biggr]
\ea
with the parameters defined in (\ref{Abbreviations}).
The $i^{\mbox{\scriptsize{th}}}$ term in the sum on the right hand side
can be identified as time evolution of a one--hole function
(\ref{FullEquations}).
In the following we choose the reaction rates according to (\ref{rates})
and rescale the $\O_n$ as in the previous section :
\be
\tilde{\O}_n(x_1,y_1,\cdots,x_n,y_n,t) \;=\;
\gamma^{\sum_{i=1}^{n}(y_i-x_i)}\,q^{\sum_{i=1}^{n}(x_i+y_i)}\,
\O_n(x_1,y_1,\cdots,x_n,y_n,t)\;.
\ee
Then the rescaled time evolution equation turns out to be invariant under
an arbitrary permutation of the variables $(x_1,\cdots,y_n)$.
This symmetry can be taken into account by introducing new coordinates
$(z_1,z_2,\cdots,z_{2n-1},z_{2n}):=(x_1,y_1,\cdots,x_n,y_n)$.
The complete system of equations then reads:\\
$\bullet$ if no holes touch the boundaries
($0<z_1, z_{2n}<L$):
\ba
&&\frac d {dt} \tilde{\O}_n(z_1,\cdots,z_{2n},t)\;=\;
-n\gamma(\gamma+\gamma^{-1})(q+q^{-1})
\tilde{\O}_n(z_1,\cdots,z_{2n},t)\label{scalnloch1}\\
&&\hspace{40mm}+\,\gamma\sum_{i=1}^{2n}\biggl[
\tilde{\O}_n(\cdots,z_i-1,z_{i+1},\cdots,t)+
\tilde{\O}_n(\cdots,z_i+1,z_{i+1},\cdots,t)\biggr]\nn
\ea
$\bullet$ if the leftmost hole touches the boundary
($0=z_1, z_{2n}<L$):
\ba
&&\frac d {dt} \tilde{\O}_n(0,z_2,\cdots,z_{2n},t)\;=\;
-((n-1)\gamma(\gamma+\gamma^{-1})(q+q^{-1})
+\gamma^2 q+q^{-1})\tilde{\O}_n(z_1,\cdots,z_{2n},t)
\label{scalnloch2}\nonumber\\
&&\hspace{25mm}+\,\gamma\sum_{i=2}^{2n}\biggl[
\tilde{\O}_n(\cdots,z_i-1,z_{i+1},\cdots,t)+
\tilde{\O}_n(\cdots,z_i+1,z_{i+1},\cdots,t)\biggr]
\ea
$\bullet$ if the rightmost hole touches the boundary
($0<z_1, z_{2n}=L$):
\ba
&&\frac d {dt} \tilde{\O}_n(z_1,\cdots,z_{2n-1},L,t)=
-((n-1)\gamma(\gamma+\gamma^{-1})(q+q^{-1})
+\gamma^2 q +q^{-1})\tilde{\O}_n(z_1,\cdots,z_{2n},t)
\label{scalnloch3}\nn\\
&&\hspace{25mm}+\,\gamma\sum_{i=1}^{2n-1}\biggl[
\tilde{\O}_n(\cdots,z_i-1,z_{i+1},\cdots,t)+
\tilde{\O}_n(\cdots,z_i+1,z_{i+1},\cdots,t)\biggr]
\ea
$\bullet$ if the leftmost and the rightmost holes touch the boundaries
($0=z_1, z_{2n}=L$):
\ba
&&\frac d {dt} \tilde{\O}_n(0,z_2,\cdots,z_{2n-1},L,t)\;=\;
-(n-1)\gamma(\gamma+\gamma^{-1})(q+q^{-1})\tilde{\O}_n(z_1,\cdots,z_{2n},t)
\label{scalnloch4}\\
&&\hspace{40mm}+\gamma\sum_{i=2}^{2n-1}\biggl[
\tilde{\O}_n(\cdots,z_i-1,z_{i+1},\cdots,t)+
\tilde{\O}_n(\cdots,z_i+1,z_{i+1},\cdots,t)\biggr]\nn
\ea
However, this set of linear equations
for the $n$--hole functions is not
closed since on the right hand side terms
with $z_i=z_{i+1}$ appear.
These terms have to be identified
with ($n-1$)--hole functions according to
\be
\tilde{\O}_n(z_1,\cdots,z_{2n})\;=\;
q^{2z_i}\,\tilde{\O}_{n-1}(z_1,\cdots,z_{i-1},z_{i+2},\cdots,z_{2n})
\;\;\mbox{if}\;\; z_i=z_{i+1}\label{koppl}
\ee
and therefore cause a coupling to the
lower sectors as mentioned before.
This coupling constitutes the main
difficulty in solving the equations.
A similar situation occurs for the spin correlation functions
of the Glauber model, and in that case a method to
solve the problem was developed by Bedeaux, Schuler and
Oppenheim \cite{Bedeaux}. It can be applied in the present
case, too. For that purpose we introduce new cumulant-like
functions $C_n$ with the property that
\begin{itemize}
\item[(1)]
$C_n$ obeys the differential equations (\ref{scalnloch1})-(\ref{scalnloch4})
\item[(2)]
$C_n(z_1,\cdots,z_{2n},t)=0$ if $z_i=z_{i+1}$
\end{itemize}
Following Ref. \cite{Bedeaux}, these functions are defined by
\ba
C_n(z_1,\cdots,z_{2n},t)&=& \sum_{\xi} (-1)^{|P|}\;(-1)^{(k-1)}\;(k-1)!\;
\tilde{\O}_{n_1}(z^\xi_1,\cdots,z^\xi_{2n_1},t)\label{cfun}\\
&& \;\;\;
\times \tilde{\O}_{n_2}(z^\xi_{2n_1+1},\cdots,z^\xi_{2n_1+2n_2},t)\cdots
\tilde{\O}_{n_k}(z^\xi_{2n-2n_k+1},\cdots,z^\xi_{2n},t)\;.\nn
\ea
The sum on the right hand side of (\ref{cfun}) runs over partitions $\xi$
of the numbers $z_1,\cdots,z_{2n}$ into $k$ subsets
$(z^\xi_1,\cdots,z^\xi_{2n_1}),(z^\xi_{2n_1+1},\cdots,z^\xi_{2n_1+2n_2}),
\cdots,(z^\xi_{2n-2n_k+1},\cdots,z^\xi_{2n})$, with sizes
$n_j$ where $j=1,\cdots,k$.
Within these subsets the $z^\xi_i$ are ordered according to their
magnitude.
$P$ is the permutation which transforms
$(z_1,\cdots,z_{2n})$ into $(z^\xi_1,\cdots,z^\xi_{2n})$
and $|P|$ denotes its (uniquely defined) sign.
The first two functions $C_1$ and $C_2$ read explicitly:
\ba
C_1(z_1,z_2,t) & = & \tilde{\O}_1(z_1,z_2,t)\label{C1}\\
C_2(z_1,z_2,z_3,z_4,t) & = & \tilde{\O}_2(z_1,z_2,z_3,z_4,t)
-\tilde{\O}_1(z_1,z_2,t)\tilde{\O}_1(z_3,z_4,t)\nn\\
&&-\tilde{\O}_1(z_1,z_4,t)\tilde{\O}_1(z_2,z_3,t)
+\tilde{\O}_1(z_1,z_3,t)\tilde{\O}_1(z_2,z_4,t)\label{C2}
\ea
The proof that the $C_n$ have the property $(2)$ by which the sectors
are decoupled, is completely
analogous to the one given in \cite{Bedeaux}.
Property (1) holds since each term on the right hand side of equation
(\ref{cfun}) solves the differential equation
(\ref{scalnloch1})-(\ref{scalnloch4}).
\\ \indent
However, this is only true if the conditions
$\d=0$ and $\a_R a_L=\a_L a_R$ are
satisfied so that the rescaled equations (\ref{scalnloch1})-(\ref{scalnloch4})
are symmetric in the starting points $x$
and end points $y$ of the holes.
In the general case, i.e. for arbitrary values of the parameters
$\d, \a_L, \a_R, a_L$ and $a_R$, this symmetry is lost and arbitrary products
of $\tilde{\O}$--functions are no longer solutions of the generalisation of
(\ref{scalnloch1})-(\ref{scalnloch4}).
Then the given construction of the $C$--functions fails.
It may not be so surprising that the method of Bedeaux et.al.,
which was developed for the Glauber model, works also here
only in the free fermion case.
To get an idea whether the general case is integrable or not,
one can apply Reshetikhin's criterion for integrability \cite{Reshetikhin}:
\be
\Big[\;H_n+H_{n+1},\Big[\;H_n,\;H_{n+1}\Big]\Big] = X_n-X_{n+1}
\ee
to the Hamiltonian (\ref{Hamiltonian}), where
$X_n$ is an arbitrary $4\times 4$ matrix.
We found that it leads to the same restriction on the reaction rates,
namely $\d=0$ and $\a_R a_L=\a_L a_R$.
It is known, however, that this criterion is not a necessary condition.
\\ \indent
Turning to the solution of the equations, one first observes that
they are inhomogeneous only for $n=1$.
A particular solution $C^P$ is again the empty lattice:
\be
C^P_1(z_1,z_2)=\gamma^{(z_2-z_1)} q^{	(z_2+z_1)}\,,
\;\;\;\;\;\;\;\;\;\; C^P_n=0\;\;\; \mbox{for}\;\;\;
n\geq 2\;.
\ee
For the solutions of the homogeneous system
we write
\be
C_n^H(z_1,\cdots,z_{2n},t)=e^{-\L t}\Phi^{(n)}(z_1,\cdots,z_{2n})
\ee
and are left with an eigenvalue problem as in the
previous section. As there, one finds four types of solutions
(c.f. Eqs. (\ref{PhiZeroMode})-(\ref{PhiTwoFermions})):
\ba
\Phi^{(n)}_{k_1,k_2,\cdots,k_{2n}}=\sum_{\t\eps S_{2n}}(-1)^{|\t|}
\prod_{i=1}^{2n}\sin\frac{\pi k_i}{L}z_{\t(i)}\label{nevec1}\\
\L=n\gamma(\gamma+\gamma^{-1})(q+q^{-1})
-2\gamma\sum_{i=1}^{2n}\cos\frac{\pi k_i}{L}\nonumber \\[4mm]
\Phi^{(n,L)}_{k_2,k_3,\cdots,k_{2n}}=\sum_{\t\eps S_{2n}}(-1)^{|\t|}
\biggl(h(L-z_{\t(1)})
\prod_{i=2}^{2n}\sin\frac{\pi k_i}{L}z_{\t(i)}\biggr)\label{nevec2}\\
\L=((n-1)\gamma(\gamma+\gamma^{-1})(q+q^{-1})
+\gamma^2 q+q^{-1})
-2\gamma\sum_{i=2}^{2n}\cos\frac{\pi k_i}{L}\nonumber\\[4mm]
\Phi^{(n,R)}_{k_1,k_2,\cdots,k_{2n-1}}=\sum_{\t\eps S_{2n}}(-1)^{|\t|}
\biggl(
\prod_{i=1}^{2n-1}\sin\frac{\pi k_i}{L}z_{\t(i)} g(z_{\t(2n)}) \biggr)
\label{nevec3}\\
\L=((n-1)\gamma(\gamma+\gamma^{-1})(q+q^{-1})
+\gamma^2 q +q^{-1})
-2\gamma\sum_{i=1}^{2n-1}\cos\frac{\pi k_i}{L}\nonumber \\[4mm]
\Phi^{(n,L,R)}_{k_2,k_3,\cdots,k_{2n-1}}=\sum_{\t\eps S_{2n}}(-1)^{|\t|}
\biggl(h(L-z_{\t(1)})
\prod_{i=1}^{2n-1}\sin\frac{\pi k_i}{L}z_{\t(i)}g(z_{\t(2n)}) \biggr)
\label{nevec4}\\
\L=(n-1)\gamma(\gamma+\gamma^{-1})(q+q^{-1})
-2\gamma\sum_{i=2}^{2n-1}\cos\frac{\pi k_i}{L}\nonumber
\ea
In these equations the $k_i$ are integers with
$1\leq k_1<\cdots<k_{2n} \leq L-1$.
The sum runs over all elements of the permutation group $S_{2n}$.
To understand these solutions we remark that each term in the sum of
(\ref{nevec1}-\ref{nevec4}) solves equations
(\ref{scalnloch1}-\ref{scalnloch4})
up to boundary conditions.
These are satified by taking the totaly antisymmetric combination.
In the case $n=1$ one recovers the results
of section 5.
\\ \indent
Comparing the eigenvalues $\L$ with the spectrum given in section 4 we
find that they correspond to the excitation of $2n$, $2n-1$ and
$2(n-1)$ fermions for the cases
(\ref{nevec1}),
(\ref{nevec2})-(\ref{nevec3}) and (\ref{nevec4}), respectively.
Furthermore, one has as many solutions as $n$--hole functions.
Therefore the system of eigenfunctions found above is complete
and the problem is thereby fully solved.
%
%----------------------------------------------------------------------
% STEADY STATE PROPERTIES
%----------------------------------------------------------------------
%
%
\section{Steady state properties}
\setcounter{equation}{0}
\noindent
The simplest application of the previous results is the investigation
of stationary properties.
The system has two steady states, namely a trivial one (the
empty lattice) and a nontrivial one where particles are present.
Considering the situation where at least one particle is present
(i.e.\ $\Omega_{stat}(0,L)=0)$ we have:
\begin{equation}
\label{StationaryState}
\Omega_{stat}(x,y) \;=\; 1-\gamma^{(L-y+x)}q^{(L-y-x)} \Phi_0(x,y)
\end{equation}
\noindent
Inserting $\Phi_0$ from Eq. (\ref{PhiZeroMode}) then gives
the following exact expression for the
concentration $c(j)=1-\Omega_{stat}(j-1,j)$ at
site $j$:
\begin{equation}
\label{LocalConcentration}
c(j) = \frac{1}{K} \biggl\{
\gamma^{2L} \biggl( (\gamma^2-1)+(q^2-1)\gamma^2(q\gamma)^{-2j} \biggr) -
q^{2L}\biggl((\gamma^2-1)q^{2-4j}+(q^2-1)(q/\gamma)^{-2j}\biggr)
\biggr\}
\end{equation}
where
\begin{equation}
K = \gamma^2\,(\gamma^{2L} + \gamma^{-2L}-q^{2L}-q^{-2L})
\end{equation}
\noindent
The density profile $c(j)$ is shown in Fig. 1 for $100$ sites,
a fixed asymmetry $q=1.2$ towards the left
and various decoagulation rates $\Delta$. As can be seen, there
are three types of curves: If $\Delta$ is small, particles are found
only near the preferred (left) boundary, while for large $\Delta$
a plateau exists together with some additional
boundary effects.
At the border between these two cases,
the concentration decays linearly in the bulk.
Computing the leading order of $c(j)$
for $L \rightarrow \infty$ gives the following explicit forms for the
profile
\begin{equation}
\label{LeadingOrder}
c(j) \;=\; \left\{
\begin{array}{lcc}
\frac{\Delta}{1+\Delta} q^{2-4j} +
\frac{q^2-1}{1+\Delta} (\frac{q}{\gamma})^{-2j} \;\;\;
& \mbox{if} & \Delta < q^2-1 \\[2mm]
\frac{\Delta}{1+\Delta} (q^{2-4j}+1-\frac{j}{L})
& \mbox{if} & \Delta = q^2-1 \\[2mm]
\frac{\Delta}{1+\Delta} + (q^2-1)
\big((q \gamma)^{-2j}-(\frac{q}{\gamma})^{2L-2j} \big)\;\;\;
& \mbox{if} & \Delta > q^2-1
\end{array} \right.
\end{equation}
\noindent From these expressions one
sees that three different inverse length
scales given by $4\ln(q)$, $2\ln(q/\gamma)$ and $2\ln(q\gamma)$ appear
which are related to those found in the fermion eigenfunctions
of section 4.
The mean concentration per site
$\overline{c}$ in leading order for $L \rightarrow \infty$ is
\begin{equation}
\label{MeanConcentration}
\overline{c} \;=\; \left\{
\begin{array}{ccc}
\frac{A}{L} & \mbox{if} & \Delta < q^2-1 \\[2mm]
\frac12\frac{\Delta}{1+\Delta} & \mbox{if} & \Delta = q^2-1  \\[2mm]
\frac{\Delta}{1+\Delta} & \mbox{if} & \Delta > q^2-1
\end{array} \right.
\end{equation}
\noindent
where
\begin{equation}
A \;=\; \frac{(1+\Delta-q^4)(1+\Delta-q^{-2})}
{(1+\Delta-q^2)(1+\Delta)(q^2-q^{-2})} \;\geq\; 1\,.
\end{equation}
\noindent
Thus we have to distinguish two different phases.
In the {\it low-density phase} $\Delta < q^2-1$ the
asymmetric diffusion is strong enough to move
the bulk particles to the boundaries where they can coagulate.
Here one has a stationary state with a finite number of particles
$A \geq 1$ and therefore the mean concentration
$\overline{c}$ is of order $1/L$.
In the {\it high-density phase} $\Delta > q^2-1$,
the bulk concentration $\frac{\Delta}{1+\Delta}$
is finite and independent of the asymmetry $q$ so that
$\overline{c}$ is of order one. At the transition
$\Delta=q^2-1$ (i.e. $q=\gamma$ or $q=\gamma^{-1}$),
one of the lengths diverges and
one has a linear decay of the concentration in the bulk.
This is also the point where the gap in the spectrum
vanishes (c.f. section 4).
\\ \indent
This phase structure also shows up in the stationary two-point
correlation function.
Denoting by $n_i$ and $n_j$ the occupation numbers at the positions
$i<j$, its connected part is given by
\ba
\label{TwoPointFunction}
g(i,j) &=&
\left\langle n_i n_j\right\rangle-
\left\langle n_i\right\rangle\left\langle n_j\right\rangle\\
&=& \left\langle (1-n_i)(1-n_j)\right\rangle-
\left\langle (1-n_i)\right\rangle\left\langle (1-n_j)\right\rangle\nonumber\\
&=&\O^{stat}_2(i-1,i,j-1,j)-\O^{stat}_1(i-1,i)\O^{stat}_1(j-1,j)\nonumber\\
&=&\gamma^{2(j-i-1)}\biggl(\O^{stat}_1(i-1,j)\O^{stat}_1(i,j-1)-
\O^{stat}_1(i-1,j-1)\O^{stat}_1(i,j)\biggr)\nonumber
\ea
\noindent
To derive the last formula one uses Eqs. (\ref{C1}), (\ref{C2}) and the f
act that for
stationary states all functions $C_n$ for $n\geq 2$ vanish because their
relaxational
spectrum has no zero eigenvalues.
Inserting the expression for the funtion $\O^{stat}_1(i,j)$ one has:
\ba
g(i,j)&=\gamma^{L+j-i-3}\; q^{L-j-i+1}&\biggl[
         \Bigl(\;h(L-i+1)-(\gamma q^{-1})\;h(L-i)\Bigr)
         \Bigl((\gamma q)\;g(j-1) - g(j)\Bigr)\nonumber \\
    & & -\Bigl((\gamma q)\;h(L-j+1)-h(L-j)\Bigr)
         \Bigl(g(i-1) - (\gamma q^{-1})\;g(i)\Bigr)\biggr]\nonumber \\
    & & -c(i)c(j)
\ea
In the symmetric case ($q=1$) the system has a ground state of product
form.
Therefore the particle concentration ${\Delta}/({1+\Delta})$
equals the mean field value and
the connected correlation function is zero. If there is an asymmetry
to the left ($q>1$), one finds the following properties
\begin{itemize}
\item In the low-density phase $\Delta<q^2-1$ the only region where
		correlations are present is the left boundary. Here the
		correlation function is negative since the dominating
		coagulation process reduces the probability that particles
		meet each other.
\item In the high-density phase $\Delta>q^2-1$ there are correlations
		at both boundaries. As in the previous case they are negative
		at the left boundary and vanish in the bulk.
                At the right boundary
		where the particle concentration is reduced, decoagulation is
		the dominating process which increases the probability of
		neighbouring particles.
\item At the phase transition $\Delta=q^2-1$ there is a linear decay of
		the correlation function for large distances.
		For large $L$ and $i/L$, $j/L$ fixed one has
		\be
		g(i/L,j/L)=(\frac{\Delta}{1+\Delta})^2\;(i/L)(1-j/L)
		\ee
		For short distances the correlations
		are nontrivial only near the left boundary where
		they have exponential form. For large $L$ and $i$,
                $j$ fixed
		\be
		g(i,j)=-(\frac{\Delta}{1+\Delta})^2\;
		(1\;+\;\gamma^{-4i+2})\;\gamma^{-4j+2}
		\ee
		In this case the remaining length scale $1/4\ln(\gamma)$
                enters and
		describes the boundary effect.
\end{itemize}
%
%
%----------------------------------------------------------------------
% CONCLUSION
%----------------------------------------------------------------------
%
\section{Concluding remarks}
\setcounter{equation}{0}
We have studied the combined effect of a preferred direction
and of open boundaries in a one-dimensional coagulation model.
By formulating it as a quantum spin chain, the complete
integrability was shown. The relaxation spectrum, although
obtained from a non-hermitean operator, turned out to be real,
with a gap determined by the asymmetry as well as by the
decoagulation. An underlying $S_3$-symmetry
of the Hamiltonian was found and
the relaxational modes seen in the hole functions
could be understood from the general result.
\\ \indent
The hierarchy of equations for these functions was considered
and decoupled by a method introduced previously for the
kinetic Ising model.
In this way a complete solution was achieved.
The decoupling approach lead to the same restrictions on
the rates as Reshetikhin's integrability criterion.
There are other solvable cases where a relation to the
XXZ Hamiltonian is used \cite{Schuetz}.
Whether further integrable situations exist, remains
open.
\\ \indent
As to the physical properties,
it was found that the model has two phases with low and
high average particle density, respectively. The corresponding
density profiles and correlations were calculated. They
contain the various lengths which arise from the interplay
of decoagulation and asymmetry. The approach to the steady
state also turns out to be interesting. Calculations show that
at the transition point an initially full system first
develops a plateau at the mean field value of the density
before a slow relaxation to the linear density profile sets in.
Only at the boundary one finds a simple algebraic decay
with a $t^{-1/2}$ law.
\\ \indent
A system with simple open ends, as considered here, is not the only
possible case. More generally, one can supply and withdraw
particles at the ends. Recent work has shown that, for pure
hopping on the chain, this leads to interesting physical phenomena
(boundary induced transitions) and to new
mathematical features (matrix-product states)
\cite{Derrida,Sandow,Stinchcombe,Essler}. Since these
processes are described by single spin operators at the
ends of the chain, they can also be included in the present
model. This is currently under investigation.
Another possible direction would be to
describe the coagulation process in more detail. This, however,
would lead to models with more than two states per site which
are more difficult to solve.
\\[10mm]
\noindent
{\bf Acknowledgements}\\[2mm]
H. H. would like to thank the
Deutsche Forschungsgemeinschaft for financial support. The authors
also thank V. Rittenberg for many valuable discussions.

\vspace{15mm}
\noindent
{\bf Figure Captions:}

\noindent
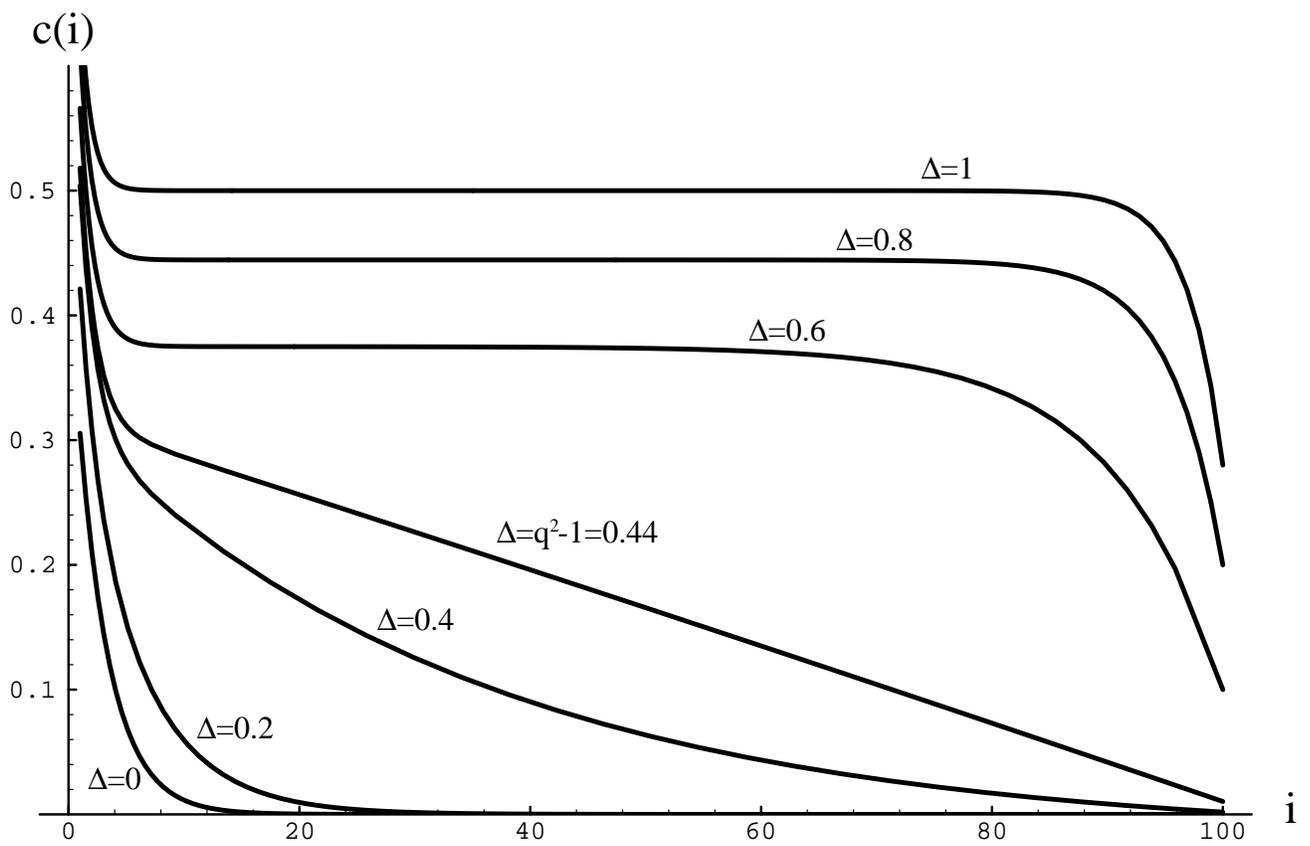
Fig. 1: Particle concentration $c(i)$ for $q=1.2$ and various
        values of $\Delta$
	on a chain of 100 sites.
\end{document}